\documentclass[10pt, twocolumn]{article}
\usepackage{amsmath,amsfonts}
\usepackage{textcomp}
\usepackage{stfloats}
\usepackage{url}
\usepackage{graphicx}
\usepackage[usenames,dvipsnames]{xcolor}
\usepackage{tikz}
\usetikzlibrary{decorations.pathreplacing, calligraphy, positioning}
\usepackage[a4paper, right = 16mm, left=16mm, top = 19mm, bottom = 30 mm]{geometry}
\usepackage[nocompress]{cite}
\usepackage{hyperref}
\hypersetup{
    colorlinks,
    linkcolor={blue!50!black},
    citecolor={blue!50!black},
    urlcolor={blue!80!black}
}

\newcommand{\freq}{{k}}
\newcommand{\lag}{u}
\newcommand{\RR}{\mathbb{R}}
\newcommand{\RRd}{{\RR^d}}
\newcommand{\freqregion}{\mathcal{K}}
\DeclareMathOperator{\sinc}{sinc}

\newcommand{\EE}[1]{\mathbb{E} \left[ #1 \right]}
\usepackage{bbm}
\newcommand{\indicator}{\mathbbm{1}}
\newcommand{\countpp}{N}
\newcommand{\de}{\mathrm{d}}
\newcommand{\var}[1]{\, {\rm var}\left( #1 \right) }
\newcommand{\norm}[1]{\left\|#1\right\|}
\newcommand{\set}[1]{\left\{#1\right\}}

\tikzstyle{figure} = [node distance = 9em]
\tikzstyle{label} = [node distance = 4.5em]
\tikzstyle{vfigure} = [figure, yshift=1.5em]

\usepackage{authblk}
\usepackage{varwidth}

\title{Visualizing the Wavenumber Content of a Point Pattern}
\date{}
\author[1]{Jake P. Grainger}
\author[2]{Tuomas A. Rajala}
\author[3]{David J. Murrell}
\author[1]{Sofia C. Olhede}
\affil[1]{Institute of Mathematics, EPFL, 1015 Lausanne, Switzerland.}
\affil[2]{Natural Resources Institute Finland, 00790 Helsinki, Finland.}
\affil[3]{\protect\begin{varwidth}[t]{\linewidth}\protect\centering Research Department of Genetics, Evolution and Environment, \par Centre for Biodiversity and Environment Research, \par University College London, UK. \protect\end{varwidth}}

\usepackage[runin]{abstract}
\usepackage[font=footnotesize,labelfont=bf]{caption}
\usepackage[tiny]{titlesec}
\usepackage{mathpazo}
\begin{document}
\maketitle

\begin{abstract}
Spatial point patterns are a commonly recorded form of data in ecology, medicine, astronomy, criminology, epidemiology and many other application fields.
One way to understand their second order dependence structure is via their spectral density function.
However, unlike time series analysis, for point patterns such approaches are currently underutilized.
In part, this is because the interpretation of the spectral representation of the underlying point processes is challenging.
In this paper, we demonstrate how to band-pass filter point patterns, thus enabling us to explore the spectral representation of point patterns in space by isolating the signal corresponding to certain sets of wavenumbers.
\end{abstract}

\section{Introduction}
Spectral analysis of time series and random fields is a very mature research area~\cite{koopmans1995spectral,percival1993spectral}. 
Spatial point patterns are found in a wide ranging set of applications, see for examples~\cite{perry2006comparison,tita2010making,reader2000using}.
In the point process setting, some work has been done to establish spectral representations \cite{daley2003introduction} and estimation of the spectral density function \cite{bartlett1964spectral, mugglestone1996practical, rajala2023what}, but the methodology has yet to be widely adopted.
In part, this is because interpreting the spectral representation of point processes is much harder than interpreting the spectral representation of random fields or time series, a problem we seek to address in this paper.

One way to think about the spectral density function is that it describes how the variability of the process can be explained by phenomena occurring at different wavenumbers.
In order to better understand such phenomena, it can be useful to isolate the spatial variability associated with a certain selection of wavenumbers and relate this to the features of the point process that this selection represent. 
In a time series context, when these wavenumbers are an interval, this projection is called a band-pass, low-pass or high-pass filter, depending on the form the interval takes.
Indeed, the earliest forms of spectral estimates were constructed by first band-pass filtering a process and then computing the variance of the output \cite{koopmans1995spectral}.

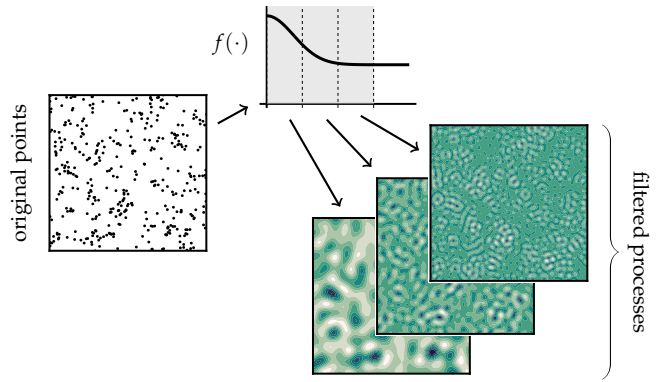
\begin{figure}
    \centering
    \begin{tikzpicture}[node distance=4em]
        \node (points) at (0,0) {\includegraphics{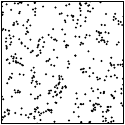}};
        \node [above right of = points, figure, xshift=1.5em, yshift=-2em] (sdf) {\includegraphics{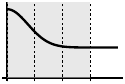}};
        \node [below of=sdf, node distance = 9em, xshift=2em] (filt1) {\includegraphics{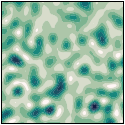}};
        \node [above of = filt1, node distance = 1.5em, xshift=2.4em] (filt2) {\includegraphics{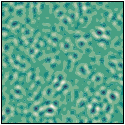}};
        \node [above of = filt2, node distance = 2em, xshift=2em] (filt3) {\includegraphics{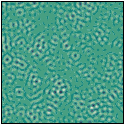}};

        \node [left of=points] (pointslabel) {\rotatebox{90}{\footnotesize original points}};
        \node [left of=sdf, yshift=0.5em] (sdflabel) {\footnotesize $f(\cdot)$};

        \node[right of=filt3, yshift= 3.3em, xshift=-0.5em] (div1) {};
        \node[right of=filt3, yshift= -7em, xshift=-0.5em] (div2) {};
        \draw [decorate, decoration = {calligraphic brace, amplitude=6pt}, pen colour={black}, black] (div1) -- node[right,xshift=0.6em]{\rotatebox{-90}{\footnotesize filtered processes}} ++ (div2);        

        \node[below of=sdf, xshift=-2em, yshift=2em] (sdf1) {};
        \node[below of=sdf, xshift=-0.75em, yshift=2em] (sdf2) {};
        \node[below of=sdf, xshift= 0.5em, yshift=2em] (sdf3) {};
        
        \draw[->,thick] (points) -- (sdf);
        \draw[->,thick] (sdf1) -- (filt1);
        \draw[->,thick] (sdf2) -- (filt2);
        \draw[->,thick] (sdf3) -- (filt3);
    \end{tikzpicture}
    \caption{Illustration of band-pass filtering of a point process. Here the points are a realization of a Thomas process, and the spectral density function (which is isotropic) is shown with a 1d slice. The filtered processes are plotted with darker shades indicating larger values, with a shared color range.}
    \label{fig:intro_example}
\end{figure}

In our case, we wish to extend such an approach of wavenumber localization to spatial point processes.
Linear filters for point processes have already been developed \cite{daley2003introduction}, and have been used for operations such as binning \cite{vere1966statistical} or studied as an inverse problem, i.e.,\ recovering points from a filter output \cite{andrieu2001bayesian}.
In contrast, we use such filtering to focus on the spatial behavior explained by different wavenumber ranges, a technique illustrated in Fig.~\ref{fig:intro_example}.

In \cite{rajala2023what} we proposed efficient methods of spectral estimation, but the interpretation of a spectral decomposition needs to be established for point processes.
This wavenumber interpretation of point processes is challenging, in part because points are essentially the roughest process one can construct, in other words, the furthest from a regular wave.
In the wavenumber domain, this corresponds to having significant variability present at all wavenumbers, in contrast to random fields, which are often assumed to be band-limited.
To address this, we show how to first project the observed point pattern into a suitable linear space, such that we retain the part of the signal associated with a given set of wavenumbers, and then, via illustrative examples, discuss how this adds insight to our understanding of such processes.

\section{Background}
We begin with some brief background on point processes.
A point process $X$ can be thought of as a random collection of locations in some space, say $\RRd$.
A formal way to characterize such a process is by the function
\begin{align}
    \countpp(B) = |X\cap B|, \quad B\subset \RRd,
\end{align}
which counts the number of points in a given region $B$ in space.
This function is called the counting measure \cite{daley2003introduction}.
A point process is said to be homogeneous if its stochastic properties are invariant to spatial shifting of the set $B$.

As in the case of time series \cite{cramer1967stationary}, homogeneous point processes have a spectral representation \cite{daley2003introduction}.
Analogously to time series, the spectral representation of a point process is usually given with the mean removed.
In particular, \cite{daley2003introduction} define the mean corrected counting measure $\countpp_0(B) = \countpp(B)-\lambda \ell(B)$, where $\lambda\ell(B) = \EE{\countpp(B)}$, for any $B$. 
Here $\lambda$ is called the intensity of the point process, and $\ell(\cdot)$ is the Lebesgue measure.
The spectral representation theorem  \cite{daley2003introduction} states
\begin{align}
    \int_{\RR^d} \phi(\lag)\countpp_0(\de \lag) &= \int_{\RR^d} \Phi(-\freq) Z(\de \freq),
\end{align}
for $\phi(\cdot)$ belonging to a certain class of functions, and where $\Phi(\freq)=\int_\RRd \phi(u) e^{-2\pi i u\cdot\freq}\de u$, $k\in\RRd$, is the Fourier transform of $\phi(\cdot)$.
We refer to $Z(\cdot)$ as the spectral process.

A consequence of homogeneity is that the spectral process is an orthogonal increment process, meaning that it is uncorrelated when evaluated over two disjoint sets.
However, the variance of the spectral process is not constant: we have traded uncorrelatedness for inhomogeneity.
Under certain conditions (see \cite{daley2003introduction}), the point process in question has a spectral density function $f(\cdot)$, which satisfies
\begin{align}
    \int_\freqregion f(\freq) \de\freq &= \var{Z(\freqregion)}, \quad \freqregion\subset \RRd.
\end{align}
In other words, the spectral density function describes the variance of the spectral process $Z(\cdot)$, and provides a decomposition of variance across scales. 

In order to understand the spectral density function, it can therefore be helpful to understand the spectral process.
Such a process cannot be visualized directly in a convenient fashion, as it is complex-valued and extremely noisy.
However, we can make a projection of the point process of interest, so that we preserve the associated spectral process at a certain limited set of wavenumbers.
This has the added benefit of producing a spatial domain process, which can be easier to interpret.
In the language of signal processing, a subset of such projections would be called band-pass filters.
Note that this is not a new way of estimating the spectral density (see our work \cite{rajala2023what}), but merely a way of separating the spectral process into different regions of the wavenumber space, and visualizing the result in the original space.

\section{Band-Pass Filtering in Space}\label{sec:bandpass}
Let $\freqregion\subset \RRd$ be a region in wavenumber space. 
Our task is to map the point process $X$ to some new process $Y(\cdot)$ whose spectral process is the same as the spectral process of $X$ in $\freqregion$, and zero outside.
Let the filtered process of $X$ be
\begin{align}
    Y(u) &= \sum_{x\in X} h_\freqregion(u-x), \quad\lag\in\RRd,
\end{align}
where $h_\freqregion(\lag)=\int_{\freqregion} e^{2\pi i \freq\cdot\lag}\de\freq$. 
In the language of signal processing, $h_\freqregion(\cdot)$ is the impulse response function.
We show in Appendix~\ref{app:proof_filter} that
\begin{align}
    Y(\lag) &= \int_\freqregion e^{2\pi i \freq\cdot\lag} Z(\de\freq) + \lambda \indicator_\freqregion(0).
\end{align}
Therefore, $Y(\cdot)$ is precisely the process we wanted.
The mean term is present when the zero wavenumber is included in the region, but is easily removed if a mean zero process is desired.

Notice that the process $Y(\cdot)$ is not a point process, but a random field.
It is possible to construct linear filters which map point processes to other point processes, see the example in \cite{brillinger1972spectral}.
However, such processes cannot be concentrated on a bounded region in wavenumber, as point processes require variance from phenomena occurring at all wavenumbers \cite{daley2003introduction}.

In practice, we must choose the region $\freqregion$. 
Whilst there are multiple choices one might consider, we prefer annuli, as they preserve the directional properties of the underlying process (see Appendix~\ref{app:region_choice} for more details).
Typically, we are interested in behavior that is not completely random, in other words when the process does not behave like a Poisson process.
Poisson processes have $f(\freq)=\lambda$ for all $\freq$ (they are the point process equivalent of white noise \cite{daley2003introduction}).
Therefore, we choose the specific wavenumber region by first estimating $f(\cdot)$, using the methodology of \cite{rajala2023what}, and then choosing an annulus in which $f(\cdot)$ is substantially different from $\lambda$.

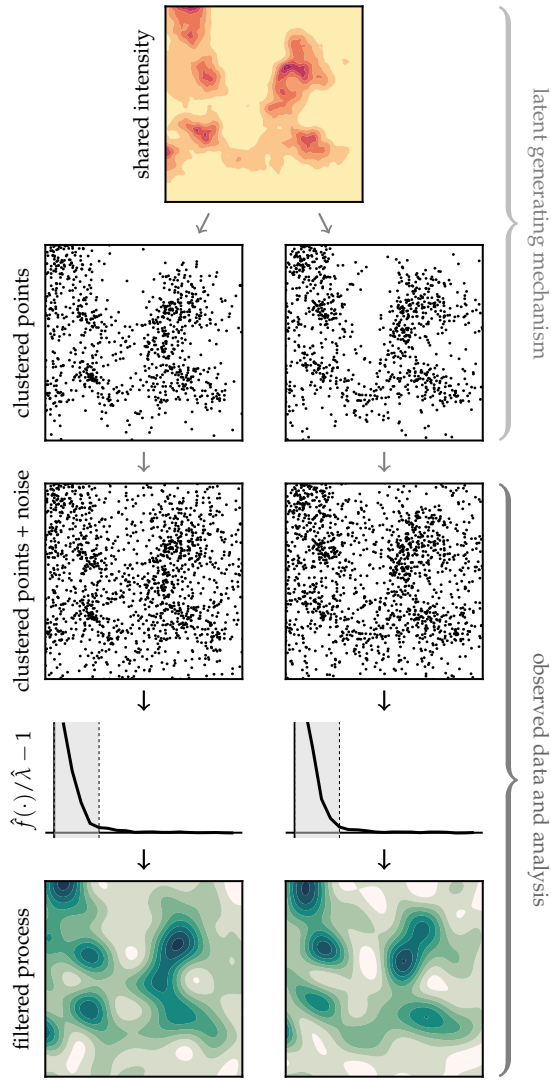
\begin{figure}[ht]
    \tikzstyle{desc_brace} = [decorate, decoration = {calligraphic brace, amplitude=6pt}, weight=heavy]
    \centering
    \begin{tikzpicture}
        \node [figure] (cp1) at (0,0) {\includegraphics{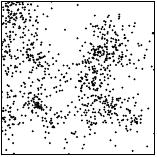}};
        \node [right of = cp1, figure] (cp2) {\includegraphics{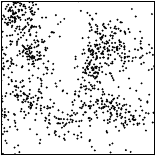}};
        \node [below of=cp1, figure] (pt1) {\includegraphics{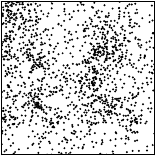}};
        \node [below of=cp2, figure] (pt2) {\includegraphics{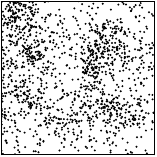}};
        \node [below of=pt1, vfigure] (sd1) {\includegraphics{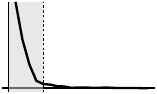}};
        \node [below of=pt2, vfigure] (sd2) {\includegraphics{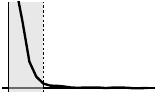}};
        \node [below of=sd1, vfigure] (fl1) {\includegraphics{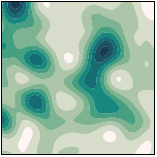}};
        \node [below of=sd2, vfigure] (fl2) {\includegraphics{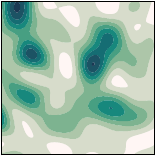}};

        \node [above of=cp1, figure, xshift = 4.5em] (intensity) {\includegraphics{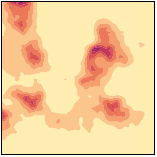}}; 

        \node [left of = intensity, label] (intlabel) {\rotatebox{90}{\footnotesize shared intensity}};
        \node [left of = cp1, label] (cplabel) {\rotatebox{90}{\footnotesize clustered points}};
        \node [left of = pt1, label] (ptlabel) {\rotatebox{90}{\footnotesize clustered points + noise}};
        \node [left of = sd1, label] (sdlabel) {\rotatebox{90}{\footnotesize $\hat f(\cdot)/\hat\lambda-1$}};
        \node [left of = fl1, label] (fllabel) {\rotatebox{90}{\footnotesize filtered process}};

        \node[right of=cp2, yshift=13em, xshift=1.5em] (div1) {};
        \node[right of=cp2, yshift=-4em, xshift=1.5em] (div2) {};
        \node[right of=pt2, yshift= 4em, xshift=1.5em] (div3) {};
        \node[right of=fl2, yshift=-4em, xshift=1.5em] (div4) {};
        \draw [desc_brace, pen colour={lightgray}, gray] (div1) -- node[right,xshift=0.6em]{\rotatebox{-90}{\footnotesize latent generating mechanism}} ++ (div2);
        \draw [desc_brace, pen colour={gray}, gray!80!black] (div3) -- node[right,xshift=0.6em]{\rotatebox{-90}{\footnotesize observed data and analysis}} ++ (div4);

        \draw[->,thick,gray] (intensity) -- (cp1);
        \draw[->,thick,gray] (intensity) -- (cp2);
        \draw[->,thick,gray] (cp1) -- (pt1);
        \draw[->,thick,gray] (cp2) -- (pt2);
        \draw[->,thick] (pt1) -- (sd1);
        \draw[->,thick] (pt2) -- (sd2);
        \draw[->,thick] (sd1) -- (fl1);
        \draw[->,thick] (sd2) -- (fl2);
    \end{tikzpicture}
    \caption{Illustration of the signal boosting obtained by low-pass filtering a point process. Darker shades indicate higher values in the contour plots, and the range is shared for the filtered processes.}
    \label{fig:sim_example}
\end{figure}

\section{Applications}
To illustrate our methodology, we explore both simulated and real data which are motivated by forest ecology.
However, point patterns are an important form of data in many different applications.
In particular, log-Gaussian Cox processes, a popular model in forest ecology \cite{ waagepetersen2016analysis}, have also been utilized to model bacteria \cite{raynaud2014spatial}, seismic activity \cite{dangelo2022local} and galaxies \cite{li2022light}, among other applications.
Furthermore, our technique can be applied to a wide range of other kinds of processes, such as Thomas processes \cite{thomas1949generalization}, hyper-uniform processes \cite{hawat2023estimating} or any other homogeneous process, such as those found in \cite{illian2008statistical}.

\subsection{Simulated Log-Gaussian Cox Processes with Noise}
We begin by considering a simulated example in which we generate two point patterns with clustering occurring in the same locations, and then add observational noise to each in the form of Poisson processes, as shown in Fig.~\ref{fig:sim_example}.
Whilst it is clear from the clustered points that there is a joint behavior between the two processes, this is much less clear when we add the noise.
However, by applying a low-pass filter, focusing on the wavenumbers at which the spectral density is substantially not $\lambda$, we can clearly see this correlation again.
To generate this example, joint clustering is obtained by generating two log-Gaussian Cox processes \cite{moller1998log} with shared intensity. 
Note that, whilst a single realization of such a process looks inhomogeneous, the random intensity is itself a homogeneous process, and thus the process as a whole is homogeneous.

\subsection{Lansing Woods}
\begin{figure}[htb]
    \centering
    \begin{tikzpicture}
        
        \node [figure] (pt1) at (0,0) {\includegraphics{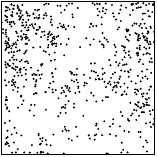}};
        \node [right of=pt1, figure] (pt2) {\includegraphics{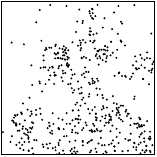}};
        
        \node [below of=pt1, vfigure] (sdf1) {\includegraphics{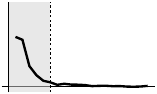}};
        \node [below of=pt2, vfigure] (sdf2) {\includegraphics{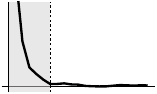}};
        
        \node [below of=sdf1, vfigure] (fl1) {\includegraphics{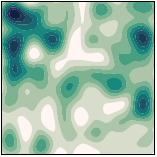}};
        \node [below of=sdf2, vfigure] (fl2) {\includegraphics{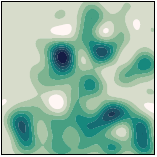}};

        \node [above of=pt1, label] (hicklab) {\footnotesize \textsc{Hickory}};
        \node [above of=pt2, label] (maplab) {\footnotesize \textsc{Maple}};
        \node [left of=pt1, label] (ptlabel) {\rotatebox{90}{\footnotesize tree locations}};
        \node [left of=sdf1, label] (sdflabel) {\rotatebox{90}{\footnotesize $\hat f(\cdot)/\hat \lambda-1$}};
        \node [left of=fl1, label] (fllabel) {\rotatebox{90}{\footnotesize filtered process}};

        \draw[->,thick] (pt1) -- (sdf1);
        \draw[->,thick] (pt2) -- (sdf2);
        \draw[->,thick] (sdf1) -- (fl1);
        \draw[->,thick] (sdf2) -- (fl2);
    \end{tikzpicture}
    \caption{The application of filtering to hickory and maple tress from the Lansing Woods data. The points, estimated spectral density function and low-pass filtered processes are shown from top to bottom respectively. In the middle plots, the shaded region delimited by the vertical line indicates the wavenumbers included in the filter. Darker shades in the contour plot indicate larger values, and the range is shared.}
    \label{fig:lansing_example}
\end{figure}
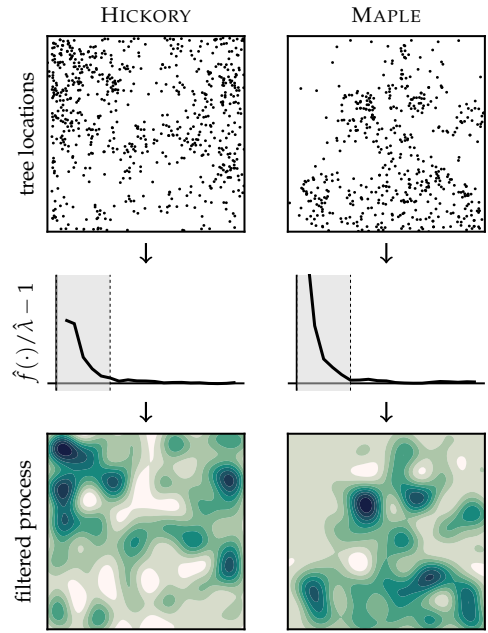

In Fig.~\ref{fig:lansing_example} we apply this filtering to a subset of the Lansing Woods data, recorded by \cite{gerrard1969new}.
For the purpose of illustration, we focus on hickories and maples, which have been shown to have a spatially segregated relationship using established point process methods \cite{wu2009summary, waagepetersen2016analysis}.
We begin by computing spectral estimates from the point patterns (using the isotropic method proposed by \cite{rajala2023what}), and notice that the spectra are substantially larger than the processes' intensities for wavenumbers below a certain threshold.
This indicates ``non-Poisson'' behavior associated with those wavenumbers.
We therefore filter both processes with a low-pass filter, designed to isolate this behavior.
We see in the filtered processes that this behavior corresponds to large scale clustering in both processes.
Furthermore, these clusters are present at different locations in space, as expected and might indicate different soil type requirements and/or direct competition between individual trees of these species.

\section{Computational Details}
The band-pass filter proposed in Section~\ref{sec:bandpass} assumes that we observe the process on the entire space $\RRd$. 
Of course this is not possible in practice, where we will be limited to observations in some bounded region $B$.
As a result, we must instead compute
\begin{align}
    Y_B(u) &=\sum_{x\in X\cap B} h_\freqregion(u-x).
\end{align}
Therefore, we are missing contributions from points outside the observational window.\footnote{In the simulated examples presented in Fig.~\ref{fig:intro_example} and Fig.~\ref{fig:sim_example}, we simulated points on a much larger domain than the one on which we evaluated the filtered process, avoiding this problem.}
However, because the impulse response functions decay fairly quickly, $Y(u)$ and $Y_B(u)$ are only different near the boundary of $B$.
Furthermore, the effect is often negligible relative to the other structures that are present.

Boundary issues withstanding, we can obtain very good properties in the wavenumber domain, obtaining near ideal band-pass filters in simulations.
In the case of time series and random fields, it is not usually possible to apply ideal band-pass filters \cite{koopmans1995spectral}.
However, in the special case of point processes we get close, as we can record the point process continuously (within a given region) without information loss, due to the sparse nature of points.
This means that, up to some edge effects, we can digitally manipulate a continuous signal, resulting in sharp attenuation in wavenumber space.

In addition, our technique requires that the process is homogeneous.
Furthermore, since it is merely a transformation of an observed realization, it is itself a stochastic process, and not an estimator converging to some true value.
As such, one must be careful when interpreting the output.
In particular, it is best used as a diagnostic and comparative tool, alongside other techniques such as spectral estimation \cite{rajala2023what}, or to aid interpretation.

\section{Conclusion}
The analysis of point patterns differs from that of other stochastic processes, such as time series and random fields.
One of these differences is that the wavenumber behavior of points needs to be present at all frequencies.
To address this here, we have developed band-pass filtering for spatial point processes.
We have shown how such an approach can be used to gain understanding of the spectral representation of a point process, exploring a variety of different models and processes of interest.
We envision that this technique will provide greater interpretability for wavenumber domain analysis techniques, enabling them to be applied to a wide range of datasets.

\section*{Data Availability}
An implementation of the methodology is available in the Julia package \verb|PointProcessFilters.jl| \cite{package}, and the R package \verb|pppgram| \cite{rajala_tuomas_2023_8268024}.
Code to generate the figures in this paper are available on GitHub \cite{jake_p_grainger_2023_8252051}, and the Lansing Woods data is available from the R package \verb|spatstat| \cite{baddeley2015spatial}.

\section*{Acknowledgments}
The work of J. P. Grainger and S. C. Olhede was supported by the European Research Council under Grant CoG 2015-682172NETS within the Seventh European Union Framework Program.

\appendix

\section{Filtering in the Wavenumber Domain}\label{app:proof_filter}
Let $\tilde Y(u) = Y(u) - \lambda \indicator_\freqregion(0)$ be the mean corrected filtered process, then
\begin{align}
        \tilde Y(u) &= \sum_{x\in X} h_\freqregion(u-x) - \lambda \indicator_\freqregion(0) \nonumber\\
        &= \int_\RRd h_\freqregion(u-s)  \countpp(\de s) - \lambda \int_\RRd h_\freqregion(u-s)\de s \nonumber\\
        &= \int_\RRd h_\freqregion(u-s) \countpp_0(\de s) \nonumber\\
        &= \int_\freqregion e^{2\pi i u\cdot \freq} Z(\de \freq)
\end{align}
as required.\footnote{
Strictly speaking, the last equality requires a more subtle argument, analogous to the proof of Corollary 4.6.VI of \cite{daley2003introduction}, and we refer the interested reader there for the technical details.}

\section{Choice of Region}\label{app:region_choice}
\begin{figure}[ht]
    \centering
    \includegraphics{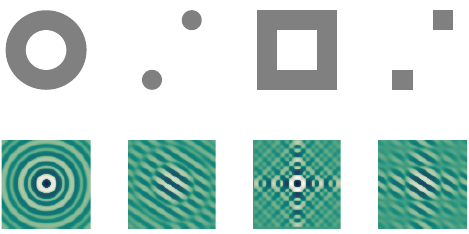}
    \caption{Some possible regions in wavenumber space which generalize a band. The transfer function is shown on the top row (where the function is either one, shown in gray, or zero, shown in white), and the corresponding impulse response on the bottom, with darker colors indicating larger values. To improve visualization, the impulse response functions are shown on a pseudo-log scale (with base 10).}
    \label{fig:transfer_function}
\end{figure}

The concept of a band in one-dimension has multiple generalizations in higher dimensions.
However, the most obvious are annuli and shifted circles, though the box equivalents may be useful in some circumstances, see Fig. \ref{fig:transfer_function}.
We prefer the first of these shapes as it is invariant to rotation and does not influence the directional properties of the original data, indeed, standard Gaussian filters  \cite{haddad1991class} used for band-pass filtering in image processing are also aiming to filter to an annulus.
The construction of $h_\freqregion(\cdot)$ for these regions is given in Appendix~\ref{app:region_const}.

As discussed above, the choice of the specific annulus should be based on regions in which the spectral density function differs substantially from $\lambda$. Often this can be chosen by eye, but one may require formal tests in some settings (e.g. some modification of \cite{mugglestone2001spectral}), though this is beyond the scope of this paper.

\section{Constructing Impulse Responses}\label{app:region_const}
In order to construct the required filters, we need only the following relations.
\subsection{Shifting}
Consider applying a symmetric shift $s$ to the region $\freqregion$, such that
\begin{equation*}
    \tilde\freqregion = (\freqregion+s) \cup (\freqregion-s) \quad\&\quad (\freqregion+s) \cap (\freqregion-s) = \emptyset
\end{equation*}
then from \cite{bateman1954tables},
\begin{equation}
    h_{\tilde\freqregion}(u) = 2\cos(2\pi s \cdot u) h_\freqregion(u).
\end{equation}
Of course, one could shift without the symmetry, but the resulting process would be complex valued in general.

\subsection{Unions}
By additivity of integrals
\begin{equation}
    h_{\freqregion_1\cup\freqregion_2}(u) = h_{\freqregion_1}(u) + h_{\freqregion_2}(u) - h_{\freqregion_1\cap\freqregion_2}(u).
\end{equation}

\subsection{Set Minus}
Again, by additivity of integrals
\begin{equation}
    h_{\freqregion_1\setminus \freqregion_2}(u) = h_{\freqregion_1}(u) - h_{\freqregion_1\cap\freqregion_2}(u).
\end{equation}

\subsection{Hyperbox}
Let $B(l)=\prod_{j=1}^d[-l_j,l_j]$, then from \cite{bateman1954tables},
\begin{equation}
    h_{B(l)}(u) = \prod_{j=1}^d 2 l_j \sinc (2 l_j u_j).
\end{equation}

\subsection{Hypersphere}
Let $S_{d}(r) = \set{k\in\RRd \;:\; \norm{k}\leq r}$, then from \cite{vembu1961fourier}
\begin{equation}
    h_{S_{d}(r)}(u) = \left(\frac{r}{\norm{u}} \right)^{d/2} \mathcal{J}_{d/2}(2\pi r \norm{u}).
\end{equation}

The annulus, cutout box, shifted sphere and shifted box are easily constructed from these operations.

\bibliography{bibliography}

\end{document}